\documentstyle[12pt,prd,epsf,aps,
amsmath,amssymb,
preprint
]
{revtex} 
\draft   
\begin{document}
\title{Slowly decaying tails of massive scalar fields
in spherically symmetric spacetimes}
\date{\today}
\author{Hiroko Koyama\thanks{Email: hiroko@allegro.phys.nagoya-u.ac.jp}
 and Akira Tomimatsu\thanks{Email: atomi@allegro.phys.nagoya-u.ac.jp}}
\address{Department of Physics, Nagoya University, Nagoya 464-8602, Japan}
\maketitle  
\begin{abstract}    
We study the dominant late-time behaviors of massive scalar fields in static
and spherically symmetric spacetimes. Considering the field evolution in the
far zone where the gravitational field is weak, we show 
under which conditions the massive field oscillates 
with an amplitude that decays slowly as $t^{-5/6}$ at very late times, as previously found in (say) the Schwarzschild case. 
Our conclusion is
that this long-lived oscillating tail is generally observed at timelike 
infinity in black hole spacetimes, while it may not be able to survive 
if the central object is a normal star. 
We also discuss that such a remarkable backscattering effect is absent 
for the field near the null cone at larger spatial distances.
\end{abstract}
\section{introduction}
One of the most 
remarkable features of wave dynamics in curved spacetimes is tails.
Scalar, electromagnetic and gravitational fields in curved spacetimes
do not, in general, propagate entirely along the null cone, but are 
accompanied by 'tails' which propagate in the interior of the null cone. 
This implies that at late times waves do not cut off
sharply but rather die off in tails.
  
In particular, it has been well established that the late-time evolution 
of massless scalar fields propagating in black-hole spacetimes
is dominated by an inverse power-law behavior,
as was first analyzed by  Price \cite{Price}.
In a brilliant work, Leaver \cite{Lea} demonstrated that late-time tails
 can be 
associated with the existence of a branch cut in the Green's function for the
wave propagation problem.
Gundlach et.al. \cite{Gun1} showed that power-law tails also characterize the
late-time evolution of radiative fields at future null infinity, 
while the decay rate is different from that of timelike infinity.
Furthermore, it has been shown that power-law tails
are a genuine feature of gravitational collapse \cite{Gun2,Marsa,BO}: 
Late-time tails develop even when no horizon is present in the background,
which means that power-law tails should be present 
in perturbations of stars, or after the implosion and subsequent explosion
of a massless field which does not result in black hole formation.
The existence of these tails was demonstrated in full non-linear 
numerical simulations of the spherically 
symmetric collapse of a self-gravitational massless scalar field.
Gundlach et al. \cite{Gun2} obtained the
power-law tails for a massless field in fully nonlinear simulations
at fixed $r$, Marsa and Choptuik \cite{Marsa} 
found them both at fixed $r$ and
along the event horizon, and Burko and Ori \cite{BO} at fixed $r$, the event
horizon, and future null infinity to very high numerical accuracy.

When the scalar field has a non-zero mass, 
the tail behaviors are quite different from massless ones.
For example, as is well known,
the tails exist even in Minkowski spacetimes,
which is related to the fact that different frequencies forming
a massive wave packet have different phase velocities\cite{MF}.
If the background spacetime is curved, it is expected that
interesting features peculiar to massive fields develop through 
the scattering due to the spacetime curvature.   

The important role of massive scalar fields has been revealed in 
elementary particle physics. For example, in higher-dimensional theories
the Fourier modes of a massless scalar
field behave like massive fields known as Kaluza-Klein modes,
and the recent development of Kaluza-Klein idea
(e.g., the Randall-Sundrum model \cite{RS} in the string theory),
strongly motivates us 
to understand the evolutional features due to the field mass in detail.
In addition, scalar fields are astrophysically important,
if boson stars made up of self-gravitating scalar fields
prove to be viable candidates for dark matter \cite{SS1}.
If such astrophysical  objects become unstable and collapse 
to form a black hole,  
both gravitational and scalar fields would be presumably radiated.
Further, as one example of curious effects
peculiar to massive fields, 
it has been argued that unstable quasi-normal modes can exist \cite{ZE,Det}.
Then the time evolution of massive scalar fields in curved spacetimes, 
(in particular, in black-hole spacetimes) would become an important
problem to be solved.

Recently it was pointed out that the late-time tails of massive scalar fields 
in Reissner-Nordstr\"{o}m spacetime  
are quite different from massless fields
in the existence of the intermediate late time tails 
\cite{HandP}(see also \cite{Burko}). 
If the Compton wave length $m^{-1}$ of a massive field
is much longer than the horizon radius of a black hole with the mass $M$,
namely $mM \ll 1$, 
each multiple moment $\psi$ of the field 
evolves into the oscillatory inverse power-law behavior
\begin{equation}
  \label{eq:HP}
  \psi \sim t^{\scriptscriptstyle - l-3/2}\sin(mt),
\end{equation}
at intermediate late times.
It is clear from (\ref{eq:HP}) that 
massive fields decay slower than massless ones, and 
waves with peculiar frequency $\omega$ quite close to $m$
mainly  contribute to massive tail,
while the dominant contribution to massless tails should be 
evaluated in the zero-frequency limit $\omega \to 0$.
Though the oscillatory power-law form (\ref{eq:HP}) 
has been numerically verified at
intermediate late times, $mM \ll mt\ll 1/(mM)^2$,
it should be  noted that the intermediate tails are
not the final asymptotic behaviors;
Another wave pattern can dominate at very late times, when 
it still remains  very difficult to determine numerically 
the exact decay rate \cite{HandP,Burko}.
In the previous paper \cite{KTR}, we have analytically found  that 
the transition from the intermediate behavior to the asymptotic  one
occurs
in nearly extreme Reissner-Nordstr\"{o}m background.
The oscillatory inverse power-law behavior of 
the dominant asymptotic tail is approximately given by 
\begin{eqnarray}
  \label{eq:asym_tail}
  \psi \sim t^{-5/6}\sin (mt),
\end{eqnarray}
independently of 
the multiple  moment $l$, and the 
decay becomes  slower than the intermediate ones.
Then, the similar result for the  decay rate has been obtained  
by considering massive scalar fields in Schwarzschild background
(in the limited cases that 
$mM\ll 1$ or $mM\gg 1$, where $M$ is the black-hole mass) \cite{KTS}
 and massive Dirac fields in Kerr-Newman backgrounds \cite{FKSY}.
Asymptotic behaviors
of massive scalar fields in dilaton black-hole backgrounds 
have been  also discussed \cite{MR}.

These results given in \cite{KTR,KTS,FKSY} suggest 
that massive fields in  black hole backgrounds
decay as $t^{-5/6}$ generally at very late times.
So it is an interesting subject to study how universally  
such a  slowly decaying tail develops.
It has been numerically shown \cite{Gun2} that 
a power-law tail develops even when the collapsing massless scalar 
field fails to produce a black hole.
This is an evidence for the late-time tail to be  a direct consequence 
of wave scattering in far distant regions.
In this paper
we prove that 
the  decay law $t^{-5/6}$ of massive scalar fields
can be essentially determined 
by the analysis
in the far zone where the gravitational field is weak.
However, we can also derive the conditions for the tails with the decay rate of $t^{-5/6}$ to dominate as an asymptotic behavior.
Considering the physical interpretation of the
 conditions, we can claim that any spherically symmetric black holes generate 
the same asymptotic tails, while the conditions may not be satisfied 
if the central object is a normal star.

In Sec. \ref{sec:model} we introduce the Green's function analysis
to investigate the time evolution of a massive scalar field
in any static, spherically symmetric spacetimes.
In Sec. III we consider the approximation valid in the far zone,
and we find the conditions
for the tail with the decay rate of $t^{-5/6}$ to develop.
The final section
is devoted to discussion, which contains a comment that the tail behavior of 
$t^{-5/6}$ breaks down as the region becomes close to the light cone.
We discuss that the tail with the decay rate of $t^{-5/6}$ can develop also
in rotating black hole spacetimes.

\section{Green's function analysis}
\label{sec:model}
\subsection{Massive scalar fields in spherically symmetric
spacetimes}
We consider the evolution of a massive scalar field in a static 
spherically symmetric background
with the asymptotically flat metric given by 
\begin{eqnarray}
  \label{eq:metric}
  ds^2=-f(r)dt^2+h(r)dr^2+r^2(d\theta ^2+\sin ^2\theta d\varphi ^2).
\end{eqnarray}
Here we do not assume the metric to be a solution 
of the vacuum or electrovac Einstein equations.
The scalar field $\Phi$ with the mass $m$ satisfies the wave equation
\begin{eqnarray}
  \label{eq:K-G}
    \Box \Phi 
&=& m^2 \Phi .
\end{eqnarray}
Resolving the field into spherical harmonics
\begin{eqnarray}
\label{bunnri}
  \Phi =\frac{\psi ^l(t,r)}{r}Y_{l,m}(\theta , \varphi),
\end{eqnarray}
hereafter we omit the index $l$ of $\psi ^l$ for simplicity,
and we obtain a wave equation for each multiple moment
\begin{eqnarray}
\label{mode-radial}
\left[\frac{\partial ^2}{\partial t^2}  -
\frac{\partial ^2}{\partial r_{\ast}^2} +V(r) \right]\psi =0,
\nonumber\\
\end{eqnarray}
where $r_{\ast}$ is the Wheeler tortoise coordinate defined by
\begin{eqnarray}
  \frac{dr_{\ast}}{dr} &=& \sqrt{\frac{h}{f}},
\end{eqnarray}
and $V$ is the effective potential
\begin{eqnarray}
  V=f\left[\frac{1}{r\sqrt{fh}}\left(\sqrt{\frac{f}{h}}\right)'
+\frac{l(l+1)}{r^2}+m^2
\right].
\end{eqnarray}

The time evolution of the radial function $\psi$
is given by
\begin{eqnarray}
  \psi (r_{\ast},t) &=&\int
[G(r_{\ast},r_{\ast}';t)\psi_t(r_{\ast}',0)
+G_t(r_{\ast},r_{\ast}';t)\psi(r_{\ast}',0)]
dr_{\ast}'
\end{eqnarray}
for $t\ge 0$, where the retarded Green's function $G$ is defined as
\begin{eqnarray}
  \left[\frac{\partial ^2}{\partial t^2}
-\frac{\partial ^2}{\partial r_{\ast}^2}
+V\right]G(r_{\ast},r_{\ast}';t)&=&\delta(t)\delta(r_{\ast}-r_{\ast}').
\end{eqnarray}
The causality condition requires that
$G(r_{\ast},r_{\ast}';t)=0$ for $t\le 0$.
In order to obtain $G(r_{\ast},r_{\ast}';t)$, we use the Fourier transform
\begin{eqnarray}
  \tilde{G}(r_{\ast},r_{\ast}';\omega)
&=&\int G(r_{\ast},r_{\ast}';t)e^{i\omega t}dt,
\end{eqnarray}
which is analytic in the upper half $\omega $ plane.
The corresponding inversion formula is
\begin{eqnarray}
\label{eq:evo_radi}
  G(r_{\ast},r_{\ast}';t)&=&
-\frac{1}{2\pi}\int _{-\infty +ic}^{\infty +ic}
\tilde{G}(r_{\ast},r_{\ast}';\omega)e^{-i\omega t}d\omega ,
\end{eqnarray}
where $c$ is some positive constant.
Now the Fourier component of the Green's function
$\tilde{G}(r_{\ast},r_{\ast}';\omega)$ is expressed in terms of two
linearly independent solutions for the homogeneous equation
\begin{eqnarray}
\label{mode-radial}
\left[\frac{\partial ^2}{\partial r_{\ast}^2}
+\omega ^2
-V
\right]\tilde{\psi}_{i}=0
\qquad \qquad i=1,2.
\end{eqnarray}
The boundary condition for the basic solution $\tilde{\psi}_1 $
is that it should be well behaved  
on the event horizon if the central object is a black hole, and 
at $r=0$ otherwise.
On the other hand, the other basic basic solution 
$\tilde{\psi}_2 $ is required to be well behaved  
at spatial infinity, $r\to \infty$.
Using these two solutions, $\tilde{G}(r_{\ast},r_{\ast}';\omega)$ 
can be written by
\begin{equation}
  \tilde{G}(r_{\ast},r_{\ast}';\omega)
=\frac{1}{W(\omega)}
\left\{
  \begin{array}{l@{\quad,\quad}l}
\tilde{\psi} _1(r'_{\ast},\omega)\tilde{\psi} _2(r_{\ast},\omega)
&r_{\ast}>r'_{\ast}\\
\tilde{\psi} _1(r_{\ast},\omega)\tilde{\psi} _2(r'_{\ast},\omega)
 &r_{\ast}<r'_{\ast},
  \end{array}
\right.
\end{equation}
where $W(\omega)$ is the Wronskian defined by
\begin{eqnarray}  
  W (\omega)&=&
\tilde{\psi} _1\tilde{\psi} _{2,r_{\ast}}
-\tilde{\psi} _{1,r_{\ast}}\tilde{\psi} _2.
\end{eqnarray}

The integrand in (\ref{eq:evo_radi}) has branch points at $\omega = \pm m$.
Considering the branch points, one may change the integration path in 
(\ref{eq:evo_radi}).
First, if $r_{\ast}-r_{\ast}'>t$, the path 
is closed in the upper half of the $\omega$
plane for the integration to converge.
Since the integrand would have no singularities in the upper half plane,
we obtain $G(r_{\ast},r_{\ast}';t) =0$ according to the causality postulate.
If $r_{\ast}-r_{\ast}'<t$, on the other hand,
the path can be deformed to the curve shown in Fig. \ref{fig1}.
As will be shown later, the late-time tails are generated owing to 
the existence of
a branch cut (in $\tilde{\psi}_2$) placed along the interval 
$-m \le \omega \le m$. 

\subsection{The analysis in a region far from the gravitational source}
It has been found in the previous papers \cite{KTR,KTS,FKSY,MR}
that the oscillatory power-law tails 
of massive scalar fields whose decay rate is $t^{-5/6}$
dominate at asymptotically late times in black-hole spacetimes.
In this paper we show that the decay law
can be simply derived 
by considering wave modes only in a far distant region, 
as a generic behaviors 
observed in any black hole spacetime.

For that purpose, we assume
\begin{eqnarray}   
\label{far-region}  
  \frac{r}{M}\gg 1,
\end{eqnarray}
where $M$ is the gravitational mass of a background field.
Then, the expansion of the metric functions 
$f$ and $h$ as a power series in $M/r$ leads to
\begin{eqnarray}
  f= 1-\frac {2M}{r}+\frac {Q^2}{r^2}+O(r^{-3})
\end{eqnarray}
and
\begin{eqnarray}
  h= 1+\frac {2M'}{r}+\frac {Q^{'2}}{r^2}+O(r^{-3}),  
\end{eqnarray}      
where $M'$, $Q$ and $Q'$ are some parameters characterizing 
the background field in more details 
in addition to the gravitational mass $M$.
Expanding (\ref{mode-radial}) in the same manner
and neglecting terms of order 
$O[(M/r)^{3}]$ and higher, we obtain the approximated form 
\begin{eqnarray}
\label{mode-radial-3}
&&
\frac{\partial ^2\tilde{\psi}}{\partial r^2}
-U
\tilde{\psi}=0,
\end{eqnarray}
where 
\begin{eqnarray}
\label{eff-pot}
  U&=&(m^2-\omega ^2)
-\frac{2M\omega ^2}{r}+\frac{2M'(m^2-\omega ^2)}{r}
-\frac{\lambda ^2-\frac 14}{r^2}.
\end{eqnarray}
The coefficient $\lambda$ in (\ref{eff-pot}) depends on the multiple moment $l$
and the other parameters $M$, $M'$, $Q$ and $Q'$.
For example, in the case of Reissner-Nordstr\"{o}m background with mass $M$
and charge $Q$, we have
\begin{eqnarray}
\label{lambda-RN}
  \lambda &=&
\sqrt{\left(l+\frac{1}{2}\right)^2 
+4m^2M^2-12\omega ^2M^2-m^2Q^2 +2\omega ^2Q^2}.
\end{eqnarray}
We keep the term of order of $O(M^2/r^{-2})$ in (\ref{mode-radial-3}),
in order to confirm 
that the decay rate of asymptotic timelike tails found 
in \cite{KTR,KTS,FKSY,MR} is independent of $\lambda$.
Introducing the variable defined as
\begin{eqnarray}
  x&=& 2\varpi r, 
\end{eqnarray}
where
\begin{eqnarray}
  \varpi =\sqrt{m^2-\omega ^2},
\end{eqnarray}
(\ref{mode-radial-3}) is rewritten by
\begin{eqnarray}
\label{modeeq-far}
  \left[
\frac{d^2}{dx^2}-\frac 14 +\frac{\kappa}{x}
-\frac{\lambda ^2-1/4}{x^2}
\right]\tilde{\psi}=0,
\end{eqnarray}  
where $\kappa $ is
\begin{eqnarray}
  \kappa =\frac{Mm^2}{\varpi}
-(M+M')\varpi .
\end{eqnarray}

\section{timelike asymptotic tail of massive scalar fields}
\label{sec:late-time}
\subsection{The wave modes}
Our aim is now to show that 
the tail with the power-law decay of  $t^{-5/6}$ 
is a generic asymptotic behavior 
in black-hole spacetimes
by using the wave modes satisfying (\ref{mode-radial-3}).
One may claim that the inner boundary condition to determine $\tilde{\psi}_1$
is missed if the analysis is limited to the range (\ref{far-region}).
Hence, we treat $\tilde{\psi}_1$ as a general solution for 
(\ref{mode-radial-3}) and reveal a condition which allows the excitation of
the asymptotic power-law tail.
Fortunately we will be able to prove that such a condition is always satisfied if the event horizon exist in the background spacetime.

First, let us give $\tilde{\psi}_2$, by requiring that it damps 
exponentially
for $|\omega| <m$ and to be purely outgoing for  $|\omega| >m$
at spatial infinity. The outer boundary condition leads to
\begin{eqnarray}
\label{modeeq-far}
\tilde{\psi}_2(\varpi ,r)&=&W_{\kappa , \lambda}(x),
\end{eqnarray} 
where $W_{\kappa , \lambda}(x)$ is the Whittaker function \cite{Abramo},
and the branch of $\varpi$ is chosen to be 
\begin{eqnarray}
  \varpi
&=&    
\left\{
  \begin{array}{l@{\quad,\quad}l}
\sqrt{m^2-\omega ^2} & \omega <m,\\
-i\sqrt{\omega ^2-m ^2}& \omega >m.
  \end{array}
\right.
\end{eqnarray}
Note that 
$W_{\kappa , \lambda}(x)$  is a many-valued function of $\varpi$, and 
there is a cut in $\tilde{\psi}_2$.
The late-time tail is generated by the contribution from the branch cut 
in $\tilde{\psi}_2$,
while $\tilde{\psi}_1$ is a one-valued function of $\varpi$,
as was shown in \cite{KTR,KTS}.
This is because the late-time tail is a consequence of backscattering. 
On the other hand, we give 
$\tilde{\psi}_1$,
 using the Whittaker functions $M_{\kappa , \lambda}$
and free parameters $a$ and $b$ as follows,
\begin{eqnarray}
\label{modeeq-far-1}  
\tilde{\psi}_1=aM_{\kappa , \lambda}(x)+bM_{\kappa , -\lambda}(x),
\end{eqnarray} 
where $a$ and $b$ will be determined if the inner boundary condition for
$\tilde{\psi}_1$ is specified.
Nevertheless the relation
\begin{eqnarray}
  \tilde{\psi}_1(\varpi)&=&\tilde{\psi}_1(e^{i\pi}\varpi)
\end{eqnarray}
should be required, 
since $\tilde{\psi}_1$ is a one-valued function 
for $\varpi$.  Then, we rewrite the parameters $a$ and $b$ as 
\begin{eqnarray}
  a
&=& j\varpi ^{-1/2-\lambda}
\end{eqnarray}
and
\begin{eqnarray}
  b
&=& k\varpi ^{-1/2+\lambda},
\end{eqnarray}  
where $j$ and $k$ are some one-valued functions for  $\varpi$.
In the following we will clarify under what kind of conditions 
for $a$ and $b$ the power-law tail with the decay rate of  $t^{-5/6}$ 
asymptotically dominates.
\subsection{Branch cut integration at timelike asymptotic regions}
As was shown in \cite{HandP,KTR,KTS},   
late-time tails 
are derived  by
the integral of $\tilde{G}(r_{\ast},r_{\ast}';\omega)$ around the branch cut
in Fig. \ref{fig1}.
Using (\ref{eq:evo_radi}),
(\ref{modeeq-far}) and (\ref{modeeq-far-1}), 
the branch cut contribution to the Green's function is given by
\begin{eqnarray}
\label{tail-part}
G^C(r_{\ast},r_{\ast}';t)
&=&-\frac{1}{2\pi}
\int _{-m}^{m}
\tilde{\psi}_1(r')
\left[
\frac{\tilde{\psi}_2(r,\varpi)}{W(\varpi)}
-\frac{\tilde{\psi}_2(r,e^{i\pi}\varpi)}{W(e^{i\pi}\varpi)}
\right]
e^{-i\omega t}d\omega
\nonumber\\
&=&\frac{1}{2\pi}
\int _{-m}^{m}
\tilde{\psi}_1(r')
\left[
\frac{ap_++bp_-}{ap_+-bp_-}-\frac{aq_++bq_-}{aq_+-bq_-}
\right]
\frac{\tilde{\psi}_1(r)}{4jk\lambda }
e^{-i\omega t}d\omega ,
\end{eqnarray} 
where
\begin{eqnarray}
  p_{\pm}&=&\frac{\Gamma(\pm2\lambda)}{\Gamma(\frac 12 \pm\lambda -\kappa)}
\end{eqnarray}
and
\begin{eqnarray}
  q_{\pm}&=&\frac{\Gamma(\pm2\lambda)}{\Gamma(\frac 12 \pm\lambda +\kappa)}
e^{i\pi (\frac 12 \mp\lambda)}.
\end{eqnarray}
Note that at very late times
\begin{eqnarray}
\label{late-time}
  mt\gg 1,
\end{eqnarray} 
the rapidly oscillating term $e^{-i\omega t}$ leads to a mutual 
cancellation between the positive and the negative parts of the integrand 
(\ref{tail-part}), except the case that the other terms of the integrand 
also change rapidly with $\omega$.
In fact, it is easy to see that 
$\tilde{\psi}_1$ neither oscillates rapidly 
nor changes exponentially with $\omega$ in the region
\begin{eqnarray}  
  \label{noninf}
  \omega r \ll \kappa.
\end{eqnarray} 
Then, if $\kappa$ remains small,  
the effective contribution to the integral in (\ref{tail-part})
is claimed to be limited to the range $|\omega - m| =O(1/t)$
or equivalently $\varpi = O(\sqrt{m/t})$
(see \cite{HandP,KTR,KTS}), and the intermediate tails becomes dominant 
at late times in the range
\begin{eqnarray}
  \label{eq:time_window}
  mM \ll mt \ll \frac{1}{(mM)^2},
\end{eqnarray}
when the integral (\ref{tail-part}) 
should be estimated under the condition
\begin{eqnarray}
\label{nonback_time}
  \kappa \simeq \frac{m^2M}{\sqrt{m^2-\omega ^2}} 
= O(mM\sqrt{mt}) \ll 1.
\end{eqnarray}
As was discussed in \cite{KTR,KTS},
the small value of $\kappa$ represents that the backscattering due to 
the spacetime curvature is not effective at intermediate late times.
It is obvious that the 
intermediate tails given by (\ref{eq:HP}) dominate
at intermediate late times (\ref{eq:time_window}),
which was numerically supported by \cite{HandP,Burko}.

As was also discussed in \cite{KTR,KTS}, however, 
the intermediate tails cannot be an asymptotic behavior,
and the long-term evolution 
from the intermediate behavior to the final one should occur.
The asymptotic tail becomes dominant at very late times such that 
\begin{eqnarray}
  mt \gg \frac{1}{m^2M^2},
\end{eqnarray}
when the effective contribution to the integral
(\ref{tail-part}) arises 
from the region
\begin{eqnarray}
\label{asympto_time}
  \kappa \simeq \frac{m^2M}{\sqrt{m^2-\omega ^2}} \gg 1,
\end{eqnarray}
which means the backscattering effect due to the curvature-induced potential
dominates. 
In the limit of $\kappa \to \infty$,
the term $(ap_++bp_-)/(ap_+-bp_-)$ becomes
\begin{eqnarray}
\label{integrand}
\frac{ap_++bp_-}{ap_+-bp_-}
& \to& 
\frac{\eta_+e^{i\pi\kappa} +\gamma_+e^{-i\pi\kappa}}
{\eta_-e^{-i\pi\kappa} +\gamma_-e^{i\pi\kappa}},
\end{eqnarray}
which includes very rapid oscillations as $e^{\pm i\pi\kappa}$, 
and we have
\begin{eqnarray}
  \eta_{\pm}&=&\Gamma(2\lambda)a\kappa ^{-\lambda}e^{-i\pi\lambda}\pm
\Gamma(-2\lambda)b\kappa ^{\lambda}e^{i\pi\lambda},
\end{eqnarray}
and   
\begin{eqnarray}
  \gamma_{\pm}&=&\Gamma(2\lambda)a\kappa ^{-\lambda}e^{i\pi\lambda}\pm
\Gamma(-2\lambda)b\kappa ^{\lambda}e^{-i\pi\lambda}.
\end{eqnarray}
Such rapidly oscillatory behaviors 
are not seen in the other term
$ (aq_++bq_-)/(aq_+-bq_-)$, which is given by
\begin{eqnarray}
 \frac{aq_++bq_-}{aq_+-bq_-}
&\to&
\frac{
a\Gamma(2\lambda)\kappa ^{-\lambda}e^{-i\pi\lambda}
+b\Gamma(-2\lambda)\kappa ^{\lambda}e^{i\pi\lambda}
}  
{
a\Gamma(2\lambda)\kappa ^{-\lambda}e^{-i\pi\lambda}
-b\Gamma(-2\lambda)\kappa ^{\lambda}e^{i\pi\lambda}
}
\end{eqnarray} 
even in the limit of $\kappa \to \infty$. 

Now we revisit the procedure through which the branch cut integration 
(\ref{tail-part}) leads to the tail with the decay rate of $t^{-5/6}$,
as was shown in \cite{KTR,KTS}.
For example, in the case of small mass field ($mM\ll 1$)
in a Schwarzschild background with mass $M$, 
$\eta _{\pm}$ and $\gamma _{\pm}$ become
\begin{eqnarray}
\eta _{\pm}&=&
\frac{\Gamma(2\lambda)^2\Gamma(1- 4i\omega M)}
{\Gamma(1/2+\lambda -2i\omega M)^2}
(4m^2M^2)^{-\lambda}e^{-i\pi\lambda}
\pm 
\frac{\Gamma(-2\lambda)^2\Gamma(1- 4i\omega M)}
{\Gamma(1/2-\lambda -2i\omega M)^2}
(4m^2M^2)^{\lambda}e^{i\pi\lambda}
\end{eqnarray}
and 
\begin{eqnarray}
\gamma _{\pm}&=&
\frac{\Gamma(2\lambda)^2\Gamma(1- 4i\omega M)}
{\Gamma(1/2+\lambda -2i\omega M)^2}
(4m^2M^2)^{-\lambda}e^{i\pi\lambda}
\pm 
\frac{\Gamma(-2\lambda)^2\Gamma(1- 4i\omega M)}
{\Gamma(1/2-\lambda -2i\omega M)^2}
(4m^2M^2)^{\lambda}e^{-i\pi\lambda}
\end{eqnarray}
in the limit of $\kappa \to \infty$ 
(see (39) and (40) in \cite{KTS}). Then the following inequalities
\begin{eqnarray}
\label{condition_saddle}  
  |\eta _{\pm}|\gtrless |\gamma _{\pm}|
\end{eqnarray} 
are satisfied when $\omega \gtrless 0$ respectively 
(see  (43) in \cite{KTS}).
When the inequalities (\ref{condition_saddle}) are hold,
the term $(ap_++bp_-)/(ap_+-bp_-)$ can be expressed as the product
of the rapidly oscillation term $e^{\pm 2i\pi\kappa}$ by 
$e^{i\varphi}$, where $\varphi$ remains in the range $0\le\varphi< 2\pi $
even if $\kappa$ becomes very large. 
The integrand (\ref{tail-part}) includes rapidly oscillating terms of 
$e^{2i\pi\kappa}$ and $e^{-i\omega t}$, 
which means physically that the scalar waves have multiple phases 
owing to the 
backscattering by the spacetime curvature, and the contribution from
these waves are canceled
 by those with the inverse phase,
unless the phase of oscillation becomes 
stationary, i.e.,
\begin{eqnarray}
\label{deq:tlsaddle}
  \frac{d}{d\omega}\left(\omega t \mp  2\pi \kappa \right)=0
\end{eqnarray}
for $\omega \gtrless 0$ respectively.
We denote $\omega$ satisfying (\ref{deq:tlsaddle}) by $\omega _0$.
Then, 
particular waves with the frequency $\omega _0$ remain without cancellation,
and contribute dominantly to the tail behaviors (see Fig. \ref{fig2}).  
In the limit of  $|\omega| \to m$ 
we obtain the solutions of (\ref{deq:tlsaddle}) as
\begin{eqnarray}
\label{sol-saddle-omega}
  t
&\simeq&
\pm \frac{2\pi\omega_0 m^2M}{(m^2-\omega _0^2)^{3/2}}
\end{eqnarray}
for $\omega \gtrless 0$ respectively, or equivalently
\begin{eqnarray}
\label{sol-saddle}
  \varpi _0
&\equiv&\sqrt{m^2-\omega _0^2}
\simeq
 m\left(\frac{2\pi M}{t}\right)^{1/3}.
\end{eqnarray}
Approximating the integration (\ref{tail-part}) by the contribution from
the close vicinity of $\omega _0$, we obtain
\begin{eqnarray}
\label{kekka:tl-tail}
  G^C(r_{\ast},r_{\ast}';t)  
&\simeq&
\frac{m}{4\sqrt{3}\lambda jk}(2\pi)^{5/6}(mM)^{1/3}
(mt)^{-5/6}
\sin (mt +\phi)
\tilde{\psi}_1(r_{\ast},m) \tilde{\psi}_1 (r_{\ast}',m).
\end{eqnarray} 
Thus we can confirm that the decay law of $t^{-5/6}$ is a result of wave
evolution in far distant regions $r\gg M$.
We also find the existence of a phase shift $\phi$ given by
\begin{eqnarray}
  \phi &=& -\frac{3}{2}(2\pi mM)^{2/3}(mt)^{1/3}-\varphi (\varpi _0)
+\frac{3}{4}\pi ,
\end{eqnarray}
which 
modulates the basic oscillation with the period of $2\pi/m$ (see \cite{KTS}).
The multiple moment $l$ and metric component $h(r)$ 
can affect only this modulation term 
in the asymptotically late-time evolution.

Note that 
if the conditions (\ref{condition_saddle}) break down, i.e., either
\begin{eqnarray}
\label{saddle_nothing}
   |\eta _{\pm}|\lessgtr|\gamma _{\pm}|,
\end{eqnarray} 
for $\omega \gtrless 0$, respectively, or
\begin{eqnarray}
\label{saddle_nothing}
   |\eta _{\pm}|=|\gamma _{\pm}|
\end{eqnarray}  
are satisfied,
all the contribution from scalar waves will be canceled more effectively.
Then the tail with the decay rate of $t^{-5/6}$ cannot survive. 
Therefore, the  conditions (\ref{condition_saddle}) are necessary
for the tail to dominate. 
The physical implication is given in the next subsection.

\subsection{Physical interpretation of the condition for the tail generation}
It is easy to see that the inequalities (\ref{condition_saddle}) 
are satisfied for the background spacetimes discussed in 
the previous papers 
(see (63) in \cite{KTR}, and  
(43) and (74) in \cite{KTS}), in which
the tail with the decay rate of $t^{-5/6}$ can dominate
at very late times.
Now we give the physical interpretation of (\ref{condition_saddle})
to discuss what background spacetime allows 
the development of the late-time tail.

Because $\lambda ^2$ in (\ref{eff-pot}) is real,
 $\lambda$ should be either real or purely imaginary.
From the expression (\ref{lambda-RN}) in Reissner-Nordstr\"{o}m background,
we find that
the small mass ($mM \ll 1$) gives a real  $\lambda$,
while the  large mass ($mM \gg 1$) gives an imaginary $\lambda$.
The  two inequalities (\ref{condition_saddle}) for $\omega \gtrless 0$ 
can be reduced to
 \begin{eqnarray}
\label{cond-real}
\frac{i}{\lambda}(ab^{\ast}-a^{\ast}b)
&\gtrless & 0
\end{eqnarray}  
when $\lambda$ is real, and 
\begin{eqnarray}
\label{cond-imaginary}
\frac{1}{\gamma} (|b|^2-|a|^2)
&\gtrless & 0
\end{eqnarray}  
when $\lambda( =i\gamma)$ is purely imaginary, respectively.

One may claim that the  condition (\ref{condition_saddle}) changes 
according to the value of $\lambda$. 
However, we can give the  unified interpretation,
independent of the value of $\lambda$,
by paying attention to $\tilde{\psi}_1$ in the region
\begin{eqnarray}
 \label{region-wave}
\omega r \ll  \frac{1}{\kappa }
\end{eqnarray}
in addition to (\ref{noninf}).
When all of  the conditions 
(\ref{far-region}), (\ref{noninf}) and (\ref{region-wave})
are satisfied, the second term which represents the  Newtonian part
in the effective potential (\ref{eff-pot}) becomes dominant, compared with the 
other terms. In this region $\tilde{\psi}_1$ is approximated by
\begin{eqnarray}
\label{eq:psi1_far_wavy}
\tilde{\psi}_1
&\simeq&
A e^{i2\sqrt{\kappa x}}
+Be^{-i2\sqrt{\kappa x}},
\end{eqnarray}
where A and B are
\begin{eqnarray}
  A&=&
\pi ^{-1/2}\kappa ^{-1/4}x^{1/4}(2\lambda)e^{-i\pi/4}
\left\{
\kappa ^{-\lambda}a\Gamma(2\lambda)e^{-i\pi\lambda}
-\kappa ^{\lambda}b\Gamma(-2\lambda)e^{i\pi\lambda}
\right\}
\end{eqnarray}
and
\begin{eqnarray}
  B&=&
\pi ^{-1/2}\kappa ^{-1/4}x^{1/4}(2\lambda)e^{-i\pi/4}
\left\{
\kappa ^{-\lambda}a \Gamma(2\lambda)e^{i\pi\lambda}
- \kappa ^{\lambda}b\Gamma(-2\lambda)e^{-i\pi\lambda}
\right\},
\end{eqnarray}
respectively (see Fig. \ref{fig3}). In this region, 
independently of $\lambda$,
 the mode clearly shows a wave behavior with the amplitudes $|A|$ and $|B|$ 
corresponding to the outgoing and ingoing parts for  $\omega >0$,
while $|A|$ and $|B|$ correspond
 to the ingoing and outgoing parts for $\omega <0$.
The difference between $|A|^2$ and $|B|^2$ is
\begin{eqnarray}
|B|^2-|A|^2
&=&2i\lambda\pi (ab^{\ast}-a^{\ast}b).
\end{eqnarray}
when $\lambda$ is real. The conditions (\ref{cond-real}) 
under which the rapid oscillation of $e^{2i\pi\kappa}$ survives
are equivalent with the inequalities
\begin{eqnarray}
\label{condition-wavy}
  |B|\gtrless|A|
\end{eqnarray}
for $\omega \gtrless 0$ respectively.
On the other hand, when $\lambda (=i\gamma)$ is purely imaginary, we have
\begin{eqnarray}
|B|^2-|A|^2
&=&  
|2\lambda|^2|\Gamma(2\lambda)|^2
(|b|^2-|a|^2)(e^{-2i\pi\lambda}-e^{2i\pi\lambda})
\nonumber\\
&=&
8\pi \gamma(|b|^2-|a|^2),  
\end{eqnarray}  
which means  the conditions (\ref{cond-imaginary}) 
are also equivalent with the inequalities (\ref{condition-wavy}).
Therefore, it is sufficient to consider the inequalities 
(\ref{condition-wavy}) 
independently of $\lambda$, as the conditions for 
the tail with the decay rate of $t^{-5/6}$ to dominate at late times.
(\ref{condition-wavy}) 
means {\it the amplitude of ingoing waves
for $\tilde{\psi}_1$ is larger than
that of outgoing waves}, in the region where 
(\ref{far-region}), (\ref{noninf})  and (\ref{region-wave}) are all satisfied.

The origin of the slowly decaying  tail as $t^{-5/6}$
of a massive scalar field can be considered a resonance 
by cooperation between 
dispersion and backscattering.
It is a common feature when the scalar field has a nonzero mass
that in far distant regions
the effective potential (\ref{eff-pot}) is
a monotonously increasing function with $r$ 
and the radial mode shows a wave behavior.
If the central object is a black hole, the conditions 
(\ref{condition-wavy}) are surely satisfied because of the existence 
of the event horizon. 
So, we can conclude
 that this long-lived oscillating tail is
generally observed in arbitrary spherical symmetric black-hole spacetimes.

\section{discussion}
\label{sec:null-tail}
We have found that whether the tail with the decay rate of  $t^{-5/6}$
develops at very late times can be
judged relevantly by wave modes only in far distant regions.
Then, even when the central object is a rotating black hole,
only the parameters $M'$ and $\lambda$ 
in the effective potential (\ref{eff-pot}) 
will be changed in far distant regions.
Strictly, background spacetimes in this paper
are limited in the class of static and spherically symmetric.
However,
since these are not relevant to the conditions (\ref{condition-wavy}),
the same tail behaviors are expected to dominate also in Kerr spacetimes.

We compare our analytical result with
their numerical simulation \cite{HandP}.
As far as the intermediate late-time behavior is concerned,
our result agrees with \cite{HandP}.  
However they claimed 
 "SI perturbation fields decay at late times slower than
any power law" in \cite{HandP}, 
which disagree with our present result
and previous ones \cite{KTR,KTS}  that the late-time tail of 
a massive scalar field is a power law with index $-5/6$.
We believe that the integration time in \cite{HandP}
is too short to find the true
asymptotic behavior.

Now we remark that the region of spacetimes where the tail
with the decay rate of $t^{-5/6}$ dominates is limited.
This feature can be understood by considering  
the behavior of $\tilde{\psi}_1$.
In the region 
\begin{eqnarray}
\label{region-limit}
  \varpi _0r \gtrsim \kappa (\omega _0),
\end{eqnarray}
$\tilde{\psi}_1$ is reduced to
\begin{eqnarray}
  \tilde{\psi}_1 
&\simeq&
\sqrt{\frac{2}{\pi}}\lambda
\left\{
\left(\frac{\kappa}{2e\varpi}\right)^{\kappa}
e^{\varpi r-\kappa\ln r}
(\eta_-e^{i\pi\kappa}+\gamma_-e^{-i\pi\kappa})
+\left(\frac{\kappa}{2e\varpi}\right)^{-\kappa}
e^{-\varpi r+\kappa\ln r}
e^{\frac{i\pi}{2}-i\pi\kappa}\gamma_-\right\}
\end{eqnarray}
and the saddle point at $\omega =\omega _0$ (\ref{sol-saddle})
is disappeared because of the terms
of $e^{\varpi r-\kappa \ln r}$ which change exponentially.
Therefore it is obvious that the
$t^{-5/6}$ tail dominates only within the region
\begin{eqnarray}
\label{tl-region}
  r\ll M^{1/3} t^{2/3}.
\end{eqnarray}

\begin{figure}
\begin{center}  
\leavevmode
\epsfxsize=80mm
\epsfbox{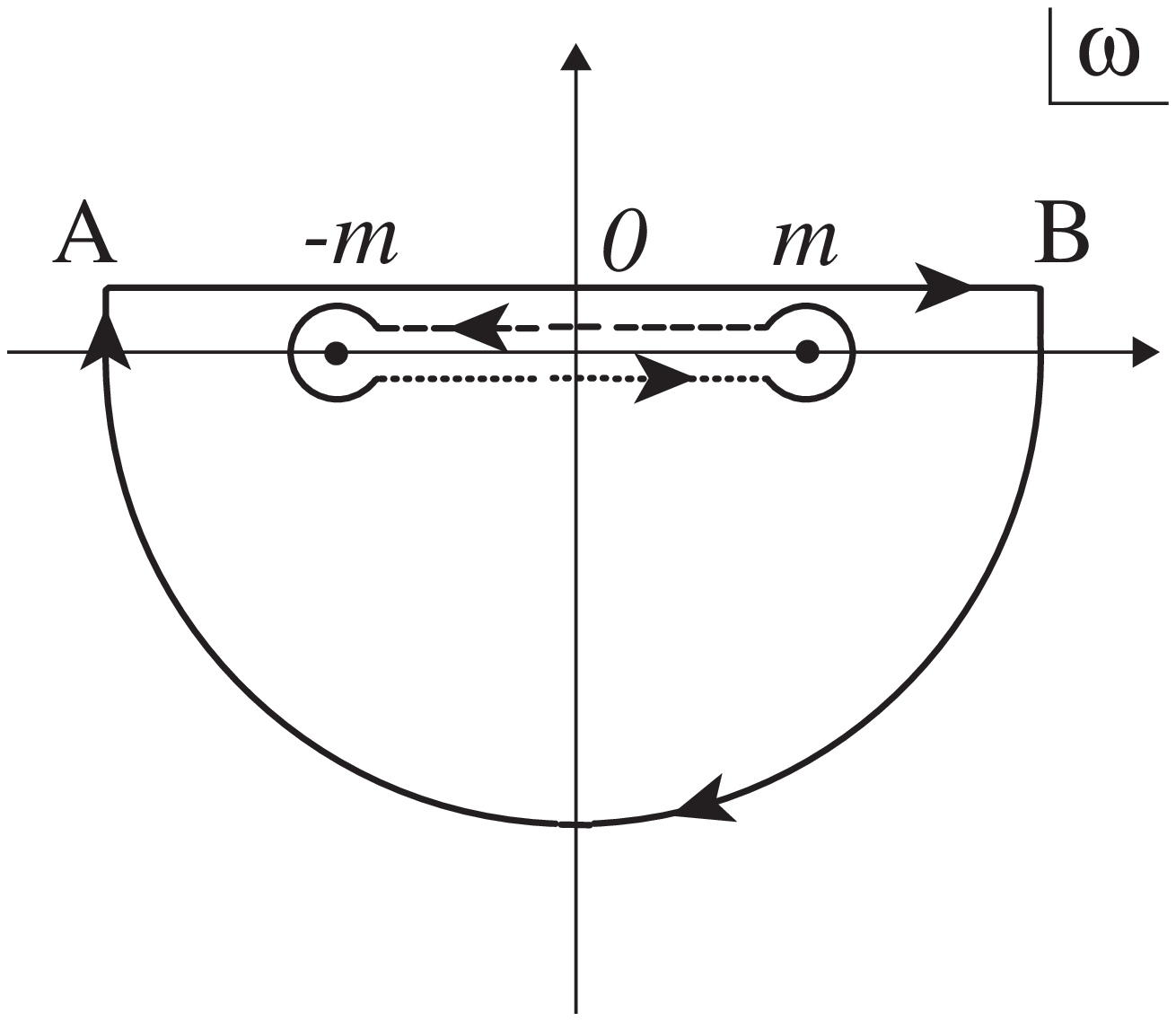}    
\caption{The integration contour for  (\ref{eq:evo_radi}),
when $t>r_{\ast}-r_{\ast}'$.
The original path corresponds to  the straight line AOB.
An integral along a branch cut placed along the interval $-m \le\omega\le m$
leads to the power-law tail.
}
\label{fig1}
\end{center}
\end{figure}

\begin{figure}
\begin{center}  
\leavevmode
\epsfxsize=80mm
\epsfbox{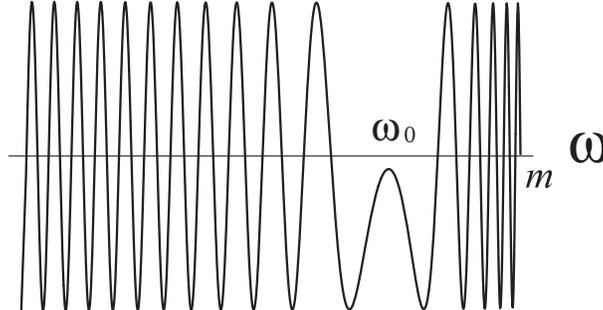}    
\caption{The schematic behaviors of the integrand 
$\tilde{G}(r_{\ast},r_{\ast}';t)e^{-i\omega t}$ 
near $\omega \simeq m$ in  (\ref{tail-part}).
The integrand includes rapidly oscillating terms of 
$e^{2i\pi\kappa}$ and $e^{-i\omega t}$, but the phase of oscillation is
stationary at $\omega =\omega _0$. }
\label{fig2}
\end{center}
\end{figure}

\begin{figure}
\begin{center}  
\leavevmode
\epsfxsize=80mm
\epsfbox{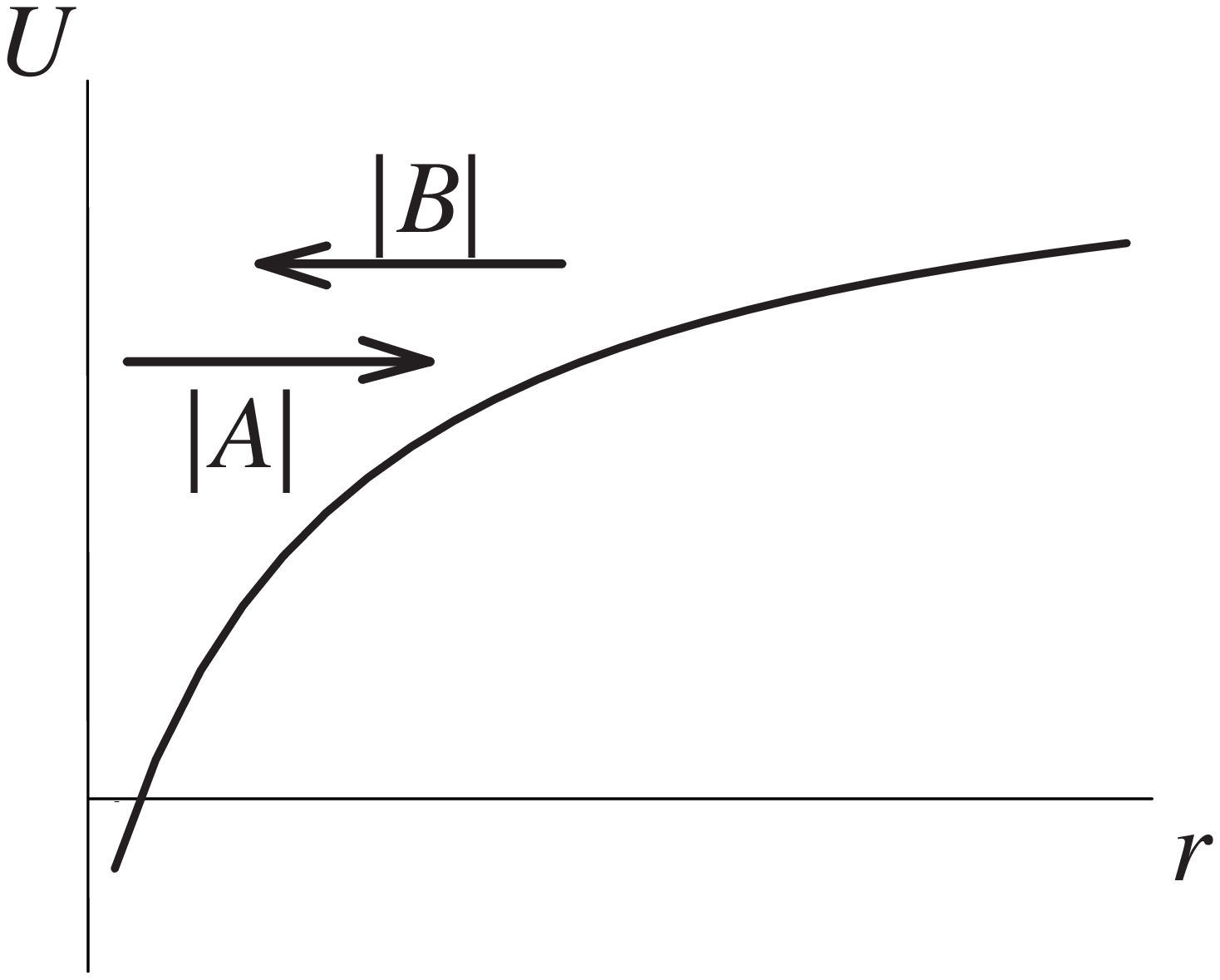}    
\caption{The schematic behaviors of the effective potential $U$
in (\ref{eff-pot}) and wave mode $\tilde{\psi}_1$ in (\ref{eq:psi1_far_wavy})
in the far region where all of the conditions 
(\ref{far-region}), (\ref{noninf}) and (\ref{region-wave})
are satisfied.}
\label{fig3}
\end{center}
\end{figure}

What kind of behaviors dominates in the region 
$r\gg M^{1/3} t^{2/3}$, in particular, near the null cone $r\to t$?
Now we must find saddle points
as solutions of the following equation
\begin{eqnarray}
\label{eq:null-saddle}
  \frac{\partial}{\partial\omega}
(-i\omega t+\varpi r -\kappa \ln r)&=&0,
\end{eqnarray}  
instead of (\ref{deq:tlsaddle}).
In general, solutions of (\ref{eq:null-saddle}) 
are complex functions of $t$ and $r$.
However  we can find a simple asymptotic solution as
\begin{eqnarray}
\label{null-saddle1}
  \omega _1&\simeq &\frac{imt}{\sqrt{t^2-\tilde{r}^2}},
\end{eqnarray}
for high frequency $\omega _1 \gg m$,  
which is compatible with the limit of 
\begin{eqnarray}
  \tilde{r} \to t,
\end{eqnarray}
where  $\tilde{r}$ is 
\begin{eqnarray}
  \tilde{r}&\equiv&
r+(M+M')\ln r,
\end{eqnarray} 
which is modified due to red shift.
The expression (\ref{null-saddle1}) means that  
we can calculate Green's function using the saddle point integration 
by deforming the integration contour into the straight line AOB 
in Fig. \ref{fig1}.  
When $\omega _1$ is a large value, 
considering the region $2\omega _1 r' \gg 1$ also,
Green's function is reduced to
\begin{eqnarray}
G(r_{\ast},r_{\ast}';t)
&\sim&
\int -\frac{e^{-i\omega t}}{2\varpi}
(e^{-\varpi R}+e^{\varpi R})d\omega ,
\end{eqnarray}
and we obtain 
\begin{eqnarray}
  \omega_1&\simeq &\frac{imt}{\sqrt{t^2-R^2}}
\end{eqnarray}
as the saddle point rather than (\ref{null-saddle1}),
where $R$ is
\begin{eqnarray}
  R&=&\tilde{r}-\tilde{r}'.
\end{eqnarray}
Approximating the integration (\ref{eq:evo_radi}) by the contribution from
the immediate vicinity of $\omega _1$, we obtain
\begin{eqnarray}
G(r_{\ast},r_{\ast}';t)
&\sim&
\Bigg|
\frac{e^{i(-\omega t+\sqrt{\omega ^2-m^2}R)}}{2i\sqrt{\omega ^2-m^2}}
\Bigg|
_{\omega =\omega _1}
\int d\omega 
\exp 
\left[
i
\Bigg|  
\frac{\partial ^2}{\partial \omega ^2}
\left(
-\omega t
+\sqrt{\omega ^2-m^2}R
\right) 
\Bigg|_{\omega =\omega _1}
(\omega -\omega _1)^2
\right]\nonumber\\
&=&
e^{i\left(-\omega _1t+\sqrt{\omega _1^2-m^2}R \right)}
2^{-3/4}m^{-1/2}(t+R)^{-1/4}(t-R)^{-1/4}.
\end{eqnarray}
Radial part of the scalar field $\Phi$ near null cone $R\simeq t$, 
together with the geometrical factor $1/r$, behaves as
\begin{eqnarray}
\label{kekka-null}
  \frac{\psi}{r} &\sim&
e^{-im(2tu)^{1/2}}
2^{-3/4}m^{-1/2}t^{-5/4}u^{-1/4},
\end{eqnarray}
where $u$ is
\begin{eqnarray}
  u&=&t-R.
\end{eqnarray}
This behavior (\ref{kekka-null}) is similar to the case of Minkowski 
spacetimes,
except for $\tilde{r}$ including red shift factor, instead of $r$.
Massive fields near null cone decay more rapidly than $t^{-5/6}$.

Finally we comment about late-time tail behaviors 
when the central object is a normal star like as a neutron star or 
a boson star.
If the expression of $\tilde{\psi}_1$ in (\ref{modeeq-far-1}) is assumed
to be extended to the region $r\le M$,
then we must require $\tilde{\psi}_1$ to be regular at $r=0$.
This leads to the equality $|A|=|B|$ 
which means that the amplitude of outgoing is equivalent with that of ingoing.
Then the tail with the decay rate of $t^{-5/6}$ never develops.
Though this extension of (\ref{modeeq-far-1}) may not be valid, we can expect the equality $|A|=|B|$ to be valid, unless some absorption of waves occurs in the inner region.
This is a future problem to be checked by giving a background gravitational field with a regular center.

\acknowledgments
The authors would like to thank Andrei V. Frolov
for valuable comments and discussions.



\begin{references}  
\bibitem{Price}R. H. Price, Phys. Rev. D {\bf 5}, 2419 (1972).
\bibitem{Lea}E W. Leaver, Phys. Rev. D {\bf 34}, 384 (1986).
\bibitem{Gun1} C. Gundlach, R H. Price, and J.Pullin,
Phys. Rev. D{\bf 49}, 883 (1994).
\bibitem{Gun2} C. Gundlach, R H. Price, and J.Pullin,
Phys. Rev. D{\bf 49}, 890 (1994).
\bibitem{Marsa}
R.L. Marsa and M.W. Choptuik,
Phys. Rev. D{\bf 54}, 4929 (1996).
\bibitem{BO}
L.M. Burko and A. Ori,
Phys. Rev. D{\bf 56}, 7820 (1997). 
\bibitem{MF}P.M. Morse and H. Feshbach, {\it Methods of Theoretical Physics}
(McGraw-Hill, New York, 1953).
\bibitem{RS}L. Randall and R. Sundrum, Phys. Rev. Lett. {\bf 83}, 4690 (1999).
\bibitem{SS1}    
E. Seidel and W-M. Suen, Phys. Rev. D{\bf 42}, 384 (1990), 
  
J. Balakrishna, E. Seidel, and W-M. Suen, Phys. Rev. D{\bf 58}, 104004 (1998).
\bibitem{ZE}T. J. Zouros and D. M. Eardley, Ann. Phys. 
{\bf 118}, 139 (1979).   
\bibitem{Det}S. Detweiler,  Phys. Rev. D{\bf 22}, 2323 (1980).
\bibitem{HandP}S. Hod and T. Piran, Phys. Rev. D {\bf 58}, 044018 (1998).
\bibitem{Burko}L. M. Burko,
{\it Abstracts of plenary talks and contributed
papers}, 15th International Conference on General Relativity and
Gravitation, Pune, 1997, p. 143, unpublished.
\bibitem{KTR}
H. Koyama and A. Tomimatsu, Phys. Rev. D {\bf 63}, 064032 (2001).
\bibitem{KTS}
H. Koyama and A. Tomimatsu, Phys. Rev. D {\bf 64}, 044014 (2001).
\bibitem{FKSY}F. Finster, N. Kamran, J. Smoller and S-T Yau,
gr-qc/0107094.
\bibitem{MR}  R. Moderski and M. Rogatko, Phys.Rev. D {\bf 64}, 044024 (2001).
\bibitem{Abramo}{\it Handbook of Mathematical Functions,} edited by 
M. Abramowitz and I.A. Stegun (Dover, New York, 1970).
\end{references}
\end{document}